\documentstyle[12pt]{article}
\begin{document}
\thispagestyle{empty}
\begin{flushright}
CERN--TH. 6602/92\\
OUT--4102--39\\
BI--TP--92/33
\end{flushright}
\vspace*{5mm}
\begin{center}
{\bf ANALYTICAL ON SHELL QED RESULTS: 3-LOOP VACUUM\\
POLARIZATION, 4-LOOP $\beta$-FUNCTION AND THE MUON ANOMALY}
\vglue 1cm
{\bf D.J. Broadhurst}
\vglue 3mm
Physics Department, Open University,\\
Milton Keynes, MK7 6AA, UK
\vglue 5mm
{\bf A.L. Kataev$^*$})
\vglue 3mm
Theory Division, CERN,\\
CH-1211 Geneva 23, Switzerland
\vglue 5mm
and
\vglue 5mm
{\bf O.V. Tarasov$^\dagger$})
\vglue 3mm
Fakult\"at f\"ur Physik, Universit\"at Bielefeld,\\
D-4800 Bielefeld 1, Germany
\vglue 1cm
{\bf Abstract}\end{center}\begin{quotation}\noindent
We present the results of analytical calculations of the 3-loop
contributions to the asymptotic photon vacuum polarization function, in the
on shell scheme, and of the 4-loop contributions to the on shell QED
$\beta$-function. These are used to evaluate various 4-loop and 5-loop
contributions to the muon anomaly. Our analytical contributions to
$(g-2)_\mu$ differ significantly from previous numerical results.
A very recent numerical re-evaluation of 4-loop muon-anomaly
contributions has yielded results much closer to ours.
\end{quotation}
\vfill
\begin{flushleft}
CERN--TH. 6602/92\\
OUT--4102--39\\
BI--TP--92/33\\
September 1992
\end{flushleft}
\vglue 5mm
\footnoterule\noindent
$*$) On leave of absence from the Institute for Nuclear Research,
117312 Moscow, Russia.\\
$\dagger$) On leave of absence from the Joint Institute for Nuclear Research,
141980 Dubna, Russia.
\newpage
\setcounter{page}{1}

{\bf 1.}~~ There are several reasons for the increase of interest in
analytical multiloop renormalization group calculations, observed during
the past decade. Firstly, the widespread use of the computer symbolic
manipulation systems SCHOONSCHIP~\cite{vel}, REDUCE~\cite{hea}, and
FORM~\cite{ver} stimulates analytical calculations in high-energy physics.
Secondly, the appearance of elegant methods~\cite{tka} for the evaluation
of massless three-loop diagrams within dimensional regularization, and
their generalization to the massive case~\cite{bro1}, have made
higher-order calculations possible. Thirdly, the detailed study of
theoretically interesting and experimentally important QED and QCD
predictions entails the consideration of the effects of high-order
radiative corrections.

Consequently, several important results have been obtained in gauge
theories, for example: the analytical 4-loop corrections to the QED
$\beta$-function in the modified minimal subtraction ($\overline{\rm MS}$)
scheme, and in the momentum (MOM) scheme (namely, the
$\Psi$-function)~\cite{gor1}; the 4-loop corrections to $\rm \sigma_{tot}\;
(e^+e^- \rightarrow hadrons)$ in QCD~\cite{gor2}; the 3-loop QED
corrections to the photon vacuum polarization function in the
$\overline{\rm MS}$ scheme~\cite{gor3}; the 2-loop corrections to the
relation between the on shell (OS) and $\overline{\rm MS}$ scheme quark
masses in QCD~\cite{gra}; the 2-loop contributions to OS renormalization
constants in QCD and QED~\cite{bro2,fle1}; the finite parts of 3-loop OS
charge renormalization~\cite{bro1}.

These results have been obtained with different computer programs, written
in different symbolic manipulation languages. Specifically, the results of
Refs.~\cite{gor1}--\cite{gor3} have been obtained with the help of the
SCHOONSCHIP program MINCER~\cite{gor4}, the calculations of
Refs.~\cite{bro1,gra,bro2} were made with the help of several REDUCE
packages, whilst the recent analysis of Ref.~\cite{fle1} was carried out
with the help of the FORM program SHELL2~\cite{fle2}. Therefore, in view of
the importance of the results obtained, it is necessary not only to study
carefully the final results, but to carry out independent cross-checks of
the systems and the computer programs involved in the above-mentioned
investigations.

In this work we combine various methods of computer symbolic
manipulation~\cite{vel}--\cite{ver}, and of evaluating multiloop Feynman
integrals~\cite{tka,bro1}, to calculate analytically the 3-loop QED
expression for the photon vacuum polarization function at large $q^2$ in
the OS scheme, the 4-loop corrections to the QED $\beta$-function in the OS
scheme, and certain 4-loop and 5-loop contributions to the muon anomalous
magnetic moment $ a_\mu = (g/2\;-\;1)_\mu$. A comparison of our analytical
results with recent numerical calculations of $a_\mu$~\cite{kin1} is
presented.

{\bf 2.}~~ We will start by deriving the basic renormalization group
relations between QED $\beta$-functions and photon vacuum polarization
functions in different renormalization schemes. This problem has been
previously analysed in Refs.~\cite{bro1,che} at the 3-loop level, and in
Refs.~\cite{gor3,kat} at the 4-loop level. Here we will apply the methods
of Ref.~\cite{bro1} to the 4-loop level.

Let us introduce some basic definitions. Firstly, we define the QED
$\beta$-function in the $\overline{\rm MS}$ scheme as
\begin{equation}
{\partial\; {\rm ln}\;\overline{\alpha}\over \partial\; {\rm ln\;\mu}}
{\Bigg |}_{\alpha_B -{\rm fixed}\atop\epsilon \rightarrow 0} =
\overline{\beta} (\overline{\alpha}) = \sum^\infty_{n=1}\overline{\beta}_n
\left({\overline{\alpha}\over \pi}\right)^n,
\end{equation}
where the renormalized $\overline{\rm MS}$ scheme coupling constant
$\overline \alpha$ is related to the bare one $\alpha_B$ by
$\overline{\alpha} (\mu)= \overline{Z}_3 (\overline{\alpha})\alpha_B,\;\mu$
is the mass unit introduced within dimensional regularization in $D = 4
-2\epsilon$ space--time dimensions, and $\overline{Z}_3
(\overline{\alpha})$ is the photon wavefunction renormalization constant in
the $\overline{\rm MS}$ scheme. The similar expression for the
$\beta$-function in the OS scheme
\vfill\eject\noindent
reads
\begin{equation}
{\partial\;{\rm ln}\;{\alpha}\over \partial\; {\rm ln}\; m} {\Bigg
|}_{\alpha_B -{\rm fixed} \atop\epsilon \rightarrow 0} = {\beta} ({\alpha})
= \sum^\infty_{n=1}{\beta}_n \left({{\alpha}\over \pi}\right)^n,
\end{equation}
where $\alpha = Z_3 \alpha_B$ is the OS scheme coupling constant (the
physical fine structure constant), $m$ is the electron pole mass, and $Z_3$
is the photon wavefunction renormalization constant in the OS
scheme\footnote{\normalsize
The $\beta$-functions and coefficients of Eqs.~(1) and~(2) are
normalized as in Refs.~\cite{bro1,kin,raf}, not as in
Refs.~\cite{gor1,gor3,kat}.}.

Using Eqs.~(1) and~(2), one can obtain the following relation:
\begin{equation}
{\overline{\beta} (\overline \alpha)\over \beta (\alpha)}{\Bigg |}_{\mu=m} =
{\partial \;{\rm ln}\; \overline{\alpha}\over \partial \;{\rm
ln}\;\alpha}{\Bigg |}_{\mu=m} = 1 +{\partial\over \partial \;{\rm ln}\;
\alpha} \left({\overline{Z}_3 (\mu = m)\over Z_3}\right).
\end{equation}
It can be shown that the ratio $\overline{\alpha} (\mu = m)/\alpha =
\overline{Z}_3 (\mu = m)/ Z_3$ has a vanishing $O(\alpha)$
contribution~\cite{bro1,che}, i.e.
\begin{equation}
{\overline\alpha (\mu=m)\over\alpha}={\overline{Z}_3 (\mu = m)\over Z_3} = 1
+ g_2 \left({\alpha\over \pi}\right)^2 + g_3 \left({\alpha\over
\pi}\right)^3 + O (\alpha^4)\; .
\end{equation}
Let us now, following the lines of Refs.~\cite{kat,fau}, expand the
coefficients $\beta_n,\; \overline{\beta}_n$, and $g_n$, in powers that
correspond to the numbers of electron loops of the corresponding diagrams
contributing to the photon vacuum polarization function. Thus we have
\begin{eqnarray}
\beta_1 &=& \beta_1^{[1]} N\;\;,\;\; \overline{\beta}_1 =
\overline{\beta}_1^{[1]} N\nonumber\\
\beta_2 &=& \beta_2^{[1]} N\;\;,\;\; \overline{\beta}_2 =
\overline{\beta}_2^{[1]} N\nonumber\\
\beta_3 &=& \beta_3^{[1]} N + \beta_3^{[2]} N^2\;\;,\;\; \overline{\beta}_3
= \overline{\beta}_3^{[1]} N + \overline{\beta}_3^{[2]} N^2\nonumber\\
\beta_4 &=& \beta_4^{[1]} N + \beta_4^{[2]} N^2 + {\beta}_4^{[3]} N^3
\nonumber\\
\overline{\beta}_4 &=& \overline{\beta}_4^{[1]} N + \overline{\beta}_4^{[2]}
N^2+ \overline{\beta}_4^{[3]} N^3\nonumber\\
g_2 &=& g_2^{[1]} N\;\;,\;\; g_3 = g_3^{[1]} N + g_3^{[2]} N^2\;,
\end{eqnarray}
where $N$ is (for formal purposes) the number of types of lepton.

Substituting Eqs.~(5) into Eq.~(4) and Eq.~(3), we get
\begin{eqnarray}
\beta_1^{[1]} &=& \overline{\beta}_1^{[1]}\;\;,\;\; \beta_2^{[1]} =
\overline{\beta}_2^{[1]}\nonumber\\
\beta_3^{[1]} &=& \overline{\beta}_3^{[1]}\;\;,\;\; \beta_3^{[2]} =
\overline{\beta}_3^{[2]} - \beta_1^{[1]} g_2^{[1]}\nonumber\\
\beta_4^{[1]} &=& \overline{\beta}_4^{[1]}\;\;,\;\; \beta_4^{[2]} =
\overline{\beta}_4^{[2]} - 2 \beta_1^{[1]} g_3^{[1]}\nonumber\\
\beta_4^{[3]} &=& \overline{\beta}_4^{[3]} - 2\beta_1^{[1]} g_3^{[2]}\;.
\end{eqnarray}
Using the concept of the QED invariant charge
\begin{equation}
\alpha_{\rm inv}(x,\alpha) = {\alpha\over 1+ \left({\alpha /
\pi}\right)\Pi\left(x,\alpha\right)} = {\overline{\alpha}\over 1+
\left({\overline\alpha /\pi}\right)
\overline{\Pi}\left(x,\overline{\alpha}\right)}\;,
\end{equation}
with $\Pi(x,\alpha) = \sum_{n=1}^\infty \Pi_n(x) \left(\alpha
/\pi\right)^{n-1},\; \overline{\Pi} (x,\overline{\alpha}) =
\sum_{n=1}^\infty \overline{\Pi}_n (x) \left(\overline{\alpha}/\pi
\right)^{n-1}$, $x = - q^2/m^2$, one can find similar relations between the
3-loop expressions for the photon vacuum polarization function in the OS
scheme, namely $\Pi(x,\alpha)$, and that in the $\overline{\rm MS}$ scheme,
renormalized at $\mu=m$,
namely $\overline{\Pi}(x,\overline{\alpha})$. Substituting the large-$x$
expansions
\begin{eqnarray}
\Pi_1 &=& (a_1 + b_1\;{\rm ln}\; x)\;\;,\;\;\overline{\Pi}_1 =
(\overline{a}_1 + \overline{b}_1 \;{\rm ln}\;x)\nonumber\\
\Pi_2 &=& (a_2 + b_2\; {\rm ln}\; x)\;\;,\;\; \overline{\Pi}_2 =
(\overline{a}_2 + \overline{b}_2 \;{\rm ln}\; x)\nonumber\\
\Pi_3 &=& (a_3 + b_3 \;{\rm ln}\; x + c_3\; {\rm ln}^2\; x)\nonumber\\
\overline{\Pi}_3 &=& (\overline{a}_3 + \overline{b}_3 \;{\rm ln}\;x
+\overline{c}_3 \;{\rm ln}^2\; x)
\end{eqnarray}
in Eq.~(7), and taking into account Eq.~(4), we get
\begin{eqnarray}
\Pi_1 &=& \overline{\Pi}_1\nonumber\\
\Pi_2 &=& \overline{\Pi}_2 - g_2\nonumber\\
\Pi_3 &=& \overline{\Pi}_3 - g_3\;.
\end{eqnarray}

This means in turn that
\begin{eqnarray}
a_1^{[1]}&=&\overline{a}_1^{[1]}\;\;,\;\; b_1^{[1]}=\overline{b}_1^{[1]}
\nonumber\\
b_2^{[1]}&=&\overline{b}_2^{[1]}\;\;,\;\; b_3^{[1]}=\overline{b}_3^{[1]}
\nonumber\\
b_3^{[2]}&=&\overline{b}_3^{[2]}\;\;,\;\; c_3^{[2]}=\overline{c}_3^{[2]}\;.
\end{eqnarray}
These identities are supported by the results of explicit calculations.
(Compare the OS scheme results, summarized in Ref.~\cite{kat}, with those
of $\overline{\rm MS}$, obtained in Ref.~\cite{gor3}.) For the 2-loop and
3-loop contributions to the constant terms of $\Pi(x,\alpha)$ and
$\overline{\Pi} (x,\overline{\alpha})$ we have
\begin{eqnarray}
a_2^{[1]} &=& \overline{a}_2^{[1]}- g_2^{[1]}\nonumber\\
a_3^{[1]} &=& \overline{a}_3^{[1]}- g_3^{[1]}\nonumber\\
a_3^{[2]} &=& \overline{a}_3^{[2]}- g_3^{[2]}\;.
\end{eqnarray}
In Eqs.~(10) and~(11) we used notations for the coefficients of the photon
vacuum polarization function similar to those defined in Eq.~(5).
Substituting Eqs.~(11) into Eqs.~(6), we get the following relations:
\begin{eqnarray}
\beta_3^{[2]} &=& \overline{\beta}_3^{[2]}-
\beta_1^{[1]}\left(\overline{a}_2^{[1]} - a_2^{[1]}\right)\nonumber\\
\beta_4^{[2]} &=& \overline{\beta}_4^{[2]}-
2\beta_1^{[1]}\left(\overline{a}_3^{[1]} - a_3^{[1]}\right)\nonumber\\
\beta_4^{[3]} &=& \overline{\beta}_4^{[3]}-
2\beta_1^{[1]}\left(\overline{a}_3^{[2]} - a_3^{[2]}\right),
\end{eqnarray}
which can be also obtained from the analysis of Ref.~\cite{kat}.

{\bf 3.}~~ Let us now consider the results of our calculations. The
analytical expressions for the 2-loop and 3-loop coefficients of the photon
vacuum polarization function in the $\overline{\rm MS}$ scheme,
$\overline{\Pi} (x,\overline{\alpha})$, were presented in Ref.~\cite{gor3}.
We are interested in the following gauge-invariant results~\cite{gor3}:
\begin{eqnarray}
\overline{a}_2^{[1]} &=& {55\over 48} - \zeta (3)\nonumber\\
\overline{a}_3^{[1]} &=& -{143\over 288} -{37\over 24} \zeta (3)
+{5\over 2}\zeta (5)\nonumber\\
\overline{a}_3^{[2]} &=& -{3701\over 2592} + {19\over 18}\zeta (3)\; .
\end{eqnarray}
They were obtained in the Feynman gauge in the process of carrying out the
work described in Ref.~\cite{gor5} with the help of the SCHOONSCHIP program
MINCER~\cite{gor4}. In the course of our work these results were confirmed
by independent calculations made in an arbitrary covariant gauge with the
help of the REDUCE program SLICER~\cite{bro3}, especially written for this
purpose.

The coefficients $g_2^{[1]},\; g_3^{[1]}$, and $g_3^{[2]}$, were calculated
in Ref.~\cite{bro1} from 2-loop and 3-loop massive propagator-like
integrals with zero external momentum, with the help of the REDUCE program
RECURSOR. They read
\begin{eqnarray}
g_2^{[1]} &=& {15\over 16}\;\;,\;\; g_3^{[1]} = {77\over 576}
+ {5\over 4}\;\zeta(2) - 2\;\zeta (2)\;{\rm ln}\;2
+ {1\over 192}\; \zeta (3) \nonumber\\
g_3^{[2]} &=& - {695\over 648} + {2\over 3}\; \zeta (2) + {7\over 64}\;
\zeta (3)\; .
\end{eqnarray}
These numbers were also confirmed in an arbitrary covariant gauge,
by means of the specially written FORM program SHELL3~\cite{tar}.

Using the results of Eqs.~(13) and~(14), we get the following expressions
for the constant contributions to the photon vacuum polarization function
in the OS scheme:
\begin{eqnarray}
a_2^{[1]} &=& {5\over 24} - \zeta (3)\nonumber\\
a_3^{[1]} &=& - {121\over 192} - {5\over 4}\;\zeta (2) + 2\;\zeta (2)\;{\rm
ln}\;2 -{99\over 64} \;\zeta (3) + {5\over 2}\; \zeta (5)\nonumber\\
a_3^{[2]} &=& - {307\over 864} - {2\over 3}\; \zeta (2) + {545\over 576} \;
\zeta (3)\; .
\end{eqnarray}
The result for $a_2^{[1]}$ is in agreement with that known for many
years~\cite{hag}. The expression for $a_3^{[2]}$ coincides with that
recently obtained in Ref.~\cite{kin}, which is known to be consistent with
numerical calculations~\cite{kin1} of 4-loop corrections to the muon
anomalous magnetic moment $a_\mu$~\cite{fau,kin}.

The result for $a_3^{[1]}$ is new. It is obtained by
combining results from Refs.~\cite{bro1,gor3} that have been exhaustively
checked by our two new programs~\cite{bro3,tar}. Using SLICER~\cite{bro3}, we
calculated the bare photon self-energy, $\Pi_{\rm B}$, to three loops at
large $-q^2$. Then we used SHELL3~\cite{tar} to calculate $\Pi_{\rm B}$ to
three loops at $q^2=0$. The constant $Z_3$ that gives the OS renormalized
result $\Pi(0,\alpha)=0$ also gave the asymptotic OS constants of Eq.~(15)
directly, without any reference to the $\overline{\rm MS}$ scheme.

Note the presence in $a_3^{[1]}$ of $\zeta (2)$ and $\zeta (2) \;{\rm ln}\;2$
terms, which are typical of calculations in the OS scheme. The origin of
the $\zeta (2) \;{\rm ln}\;2$ term is well understood: it comes from a
scalar massive integral that also contributes to the 2-loop correction to
the fermion anomalous magnetic moment~\cite{bro1,gra,fle1}. Notice also
that the $\zeta (5)$ contribution to the 3-loop expression of $\Pi
(x,\alpha)$, like the 2-loop $\zeta(3)$ term, is independent of the choice
of scheme. (Compare Eq.~(15) with Eq.~(13).) This fact can be understood by
using the methods of Ref.~\cite{tka}.

It should be stressed that our numerical value of $a_3^{[1]}$, namely
\begin{equation}
a_3^{[1]} = 0.3268745\ldots\quad ({\rm this\;work})\; ,
\end{equation}
strongly differs from the numerical result
\begin{equation}
a_3^{[1]} = 1.356(41)+O\left({m_e\over m_\mu}\right)\quad ({\rm
Ref.~\cite{kin1}})
\end{equation}
obtained from the analysis of the results of numerical calculation of the
4-loop corrections to $a_\mu$~\cite{kin1} by the methods of
Ref.~\cite{lau2}, taking into account the correct analytical expression for
the $a_3^{[2]}$ term\footnote
{\normalsize The analysis of Ref.~\cite{kin1} gives $a_3 =
a_3^{[1]} + a_3^{[2]} = 1.041(41) + O (m_e/m_\mu)$.}.

{\bf 4.}~~ By using our new analytical expression for the $a_3^{[1]}$ term,
we can also obtain the analytical expression for the set of 4-loop diagrams
contributing to $a_\mu$ formed by inserting the sum of 15 one-particle
irreducible sixth order single-electron-loop vacuum-polarization diagrams
in a second-order muon vertex. (See Fig.~1.) Indeed, using the methods of
Ref.~\cite{lau2}, which agree with those of Ref.~\cite{bar}, we get
\begin{equation}
a_\mu \;{(\rm Fig.\;1)} = \left(-a_3^{[1]} I_0 - b_3^{[1]} I_1 - 2 b_3^{[1]}
I_0\; {\rm ln}\; \left({m_\mu\over m_e}\right) + O \left({m_e\over
m_\mu}\right)\right) \left({\alpha\over \pi}\right)^4,
\end{equation}
where $I_0 = 1/2$, $I_1 = -5/4$, and $b_3^{[1]} = 1/32$~\cite{ros}.
Substituting the analytical value of $a_3^{[1]}$ from Eq.~(15), we have
\begin{eqnarray}
a_\mu \;{(\rm Fig.\;1)} &=& \left[{17\over 48}+{5\over 8}\; \zeta (2) -
\zeta (2)\;{\rm ln}\;2 + {99\over 128}\; \zeta (3)
- {5\over 4}\; \zeta (5) \right.\nonumber\\
&&\left.{} - {1\over 32}\;{\rm ln} \;\left({m_\mu\over m_e}\right)
+ O\left({m_e\over m_\mu}\right) \right] \left({\alpha\over
\pi}\right)^4\nonumber\\
&=& \left[ - 0.290987 + O \left({m_e\over m_\mu}\right)\right]
\left({\alpha\over \pi}\right)^4\quad ({\rm this\;work})
\end{eqnarray}
whilst the numerical calculations of Ref.~\cite{kin1} gave
\begin{equation}
a_\mu\;{\rm (Fig.\;1)} = - 0.7945 (202) \left({\alpha\over\pi}\right)^4\quad
({\rm Ref.~\cite{kin1}})\; .
\end{equation}
The discrepancy between Eqs.~(19) and~(20) is much greater than any
revealed by the analytical calculations in Ref.~\cite{caf} of some 8th
order contributions to the electron anomaly.

The discrepancy between Eqs.~(16) and~(17) also leads to a substantial
change to the 10th order contributions to $a_\mu$ of the diagrams of
Fig.~2. Indeed, using the methods of Ref.~\cite{lau2}, one can get the
following expression for them~\cite{kat}:
\begin{eqnarray}
a_\mu\;{\rm (Fig.\;2)} &=& \bigg [ {25\over 18}\; a_3^{[1]} - {1\over 9} -
{1\over 24} \;\zeta (2) + \left({5\over 36} - {2\over 3}\; a_3^{[1]}\right)
\;{\rm ln}\left({m_\mu\over m_e}\right)\nonumber\\
&&{} - {1\over 24}\; {\rm ln}^2 \left({m_\mu\over m_e}\right)
+O\left({m_e\over m_\mu}\right)\bigg ] \left({\alpha\over\pi}\right)^5 .
\end{eqnarray}
Taking the value of the $a_3^{[1]}$ term from Eq.~(15), we are now able to
obtain the following totally analytical expression for Eq.~(21):
\begin{eqnarray}
a_\mu\;{\rm (Fig.\;2)} &=& \bigg [ -{3409\over 3456} - {16\over 9} \;\zeta
(2) + {25\over 9}\; \zeta (2) \;{\rm ln}\;2
- {275\over 128}\; \zeta (3) + {125\over 36}\; \zeta (5)\nonumber\\
&&{}+\left({161\over 288} + {5\over 6}\; \zeta (2)
- {4\over 3}\; \zeta (2)\;{\rm ln}\; 2 + {33\over 32}\; \zeta (3)
- {5\over 3} \;\zeta (5) \right)
{\rm ln}\left({m_\mu\over m_e}\right)\nonumber\\
&&{}-{1\over 24} \;{\rm ln}^2 \left({m_\mu\over m_e}\right) + O
\left({m_e\over m_\mu}\right)\bigg ]\left({\alpha\over \pi}\right)^5.
\end{eqnarray}
The corresponding numerical result reads
\begin{equation}
a_\mu\;{\rm (Fig.\;2)} = \left[-1.3314 + O \left({m_e\over
m_\mu}\right)\right] \left({\alpha\over \pi}\right)^5 \quad ({\rm
this\;work})\; .
\end{equation}
It should be compared with the similar expression obtained in
Ref.~\cite{kat} from Eq.~(21), after taking into account the `old'
$a_3^{[1]}$ value from Eq.~(17), namely
\begin{equation}
a_\mu\;{\rm (Fig.\;2)} = \left[-3.560 (89) + O\left({m_e\over
m_\mu}\right)\right] \left({\alpha\over \pi}\right)^5 \quad ({\rm
Ref.~\cite{kat}})\; .
\end{equation}
Therefore, one way of resolving the discrepancy between Eqs.~(16) and~(17)
might be an accurate numerical calculation of the diagrams of Fig.~2.

{\bf 5.}~~ Our next step is the determination of the 4-loop correction to
the QED $\beta$-function in the OS scheme, in analytical form. We first
remark that the results of the analytical calculations of the 4-loop
corrections to the QED $\beta$-function in the $\overline{\rm MS}$ scheme,
namely to the $\overline\beta$-function defined by Eq.~(1), are~\cite{gor1}
\begin{eqnarray}
\overline{\beta}_1 &=& {2\over 3} \;N\;\;,\;\;\overline{\beta}_2 = {1\over
2}\; N\nonumber\\
\overline{\beta}_3 &=& -{1\over 16}\;N - {11\over 72}\;N^2\nonumber\\
\overline{\beta}_4 &=& -{23\over 64}\; N+ \left({95\over 432}-{13\over 18}\;
\zeta (3) \right) N^2 - {77\over 1944}\;N^3\;.
\end{eqnarray}

Using now Eqs.~(5) and~(6), and the results of Eq.~(14), we get
\begin{equation}
\beta_3 = - {1\over 16}\; N - {7\over 9}\; N^2\;.
\end{equation}
At $N=1$, Eq.~(26) coincides with the result $\beta_3=-121/144$ obtained in
Ref.~\cite{raf}.

The 4-loop coefficient $\beta_4$ has the following form:
\begin{eqnarray}
\beta_4 &=& - {23\over 64} N + \left({1\over 24} - {5\over 3}\; \zeta (2) +
{8\over 3}\; \zeta (2) \;{\rm ln}\;2 - {35\over 48} \zeta (3)\right)
N^2\nonumber\\
&&{}+ \left({901\over 648}- {8\over 9}\;\zeta (2) - {7\over 48} \;\zeta
(3)\right)N^3,
\end{eqnarray}
where the powers of $N$ serve merely to distinguish contributions with
corresponding numbers of electron loops. (With leptons of different mass,
the finite parts of OS charge renormalization involve
dilogarithms~\cite{bro1}, and the OS expansions of Eq.~(8) involve
logarithms~\cite{fau} of mass ratios; both are absent from the
$\overline{\rm MS}$ scheme.) Setting $N=1$, we obtain
\begin{equation}
\beta_4 = {5561\over 5184}-{23\over 9}\;\zeta (2) + {8\over 3} \;\zeta (2)
\;{\rm ln}\;2 -{7\over 8}\;\zeta (3)\; ,
\end{equation}
which may also be obtained from the $m$-independence of
$\Lambda^{\overline{\rm MS}}_{\rm QED}$, as given in Ref.~\cite{bro1}. The
order $N$ term in Eq.~(27) is scheme independent. The $N^3$ contribution
coincides with the result of Ref.~\cite{kin}. The $N^2$ contribution is
new. In view of the results described in Sections 3 and 4 it differs from
the one obtained previously~\cite{gor3} taking into account the `old'
numerical value of the $a_3^{[1]}$ term. (Compare Eq.~(16) with Eq.~(17).)

Let us now compare the numerical behaviour of the 4-loop approximations of
the $\beta$-functions in the $\overline{\rm MS}$ and OS schemes with the
corresponding function in the MOM scheme, namely with the $\Psi$-function,
which we here normalize in correspondence with Eqs.~(1) and~(2), as
follows:
\begin{equation}
\Psi (\alpha_{\rm MOM}) =\frac{\partial\;{\rm ln}\;\alpha_{\rm MOM}}
{\partial\;{\rm ln}\;\alpha}\beta(\alpha) =\frac{\partial\;{\rm ln}\;
\alpha_{\rm MOM}}{\partial\;{\rm ln}\;\overline{\alpha}}
\overline{\beta}(\overline{\alpha})\; ,
\end{equation}
where $\alpha_{\rm MOM}$ is the invariant charge of Eq.~(7), at fixed
$q^2=-\lambda^2$. The expressions for the functions
$\overline{\beta}(\alpha)$ and $\Psi (\alpha)$ can be obtained from the
results of Ref.~\cite{gor1}:
\begin{equation}
\overline{\beta}(\alpha) = 0.667 \left({\alpha\over \pi}\right) +
0.5 \left({\alpha\over \pi}\right)^2 - 0.215 \left({\alpha\over
\pi}\right)^3 - 1.047 \left({\alpha\over \pi}\right)^4\;...
\end{equation}
\begin{equation}
\Psi (\alpha) = 0.667 \left({\alpha\over \pi}\right) + 0.5
\left({\alpha\over \pi}\right)^2 + 0.100 \left({\alpha\over \pi}\right)^3 -
1.202 \left({\alpha\over \pi}\right)^4\;...
\end{equation}
In the OS scheme the analogous series reads
\begin{equation}
\beta (\alpha) = 0.667 \left({\alpha\over \pi}\right) + 0.5
\left({\alpha\over \pi}\right)^2 - 0.840 \left({\alpha\over \pi}\right)^3 -
1.142 \left({\alpha\over \pi}\right)^4\;...
\end{equation}

Notice that now, in all three cases, the 4-loop coefficients are negative
and have the same order of magnitude. This is a welcome result, in view of
the natural expectation that the physical conclusions should be less
scheme-dependent after taking into account higher-order perturbative
corrections.

Finally, we use the new numerical value of the 4-loop coefficient of the OS
$\beta$-function of Eq.~(32), to obtain a slight modification of the
estimate presented in Ref.~\cite{kat} for the renormalization-group
constrained~\cite{lau2} 10th order contributions to $a_\mu$. Our result
reads:
\begin{equation}
a_\mu^{(10)}({\rm R.G.})
= \left[B_5 + 82.92 (78)\right] \left({\alpha\over \pi}\right)^5
\quad ({\rm this\;work})\; ,
\end{equation}
where $B_5$ is the unknown 5-loop mass-independent contribution (expected
to be positive) and the uncertainty in the numerical value of the
renormalization-group determined mass-dependent term derives largely from
the uncertainty in $B_4=-2.503(55)$~\cite{kin1}. Our result differs little
from that of Ref.~\cite{kat}, namely $a_\mu^{(10)}({\rm R.G.}) = \left[B_5
+ 86.57 (78)\right] (\alpha/\pi)^5$, obtained using Eq.~(17).

{\bf 6.}~~ In conclusion: the 4-loop and 5-loop muon-anomaly contributions
of Eqs.~(19), (23) and~(33) follow from our new results for the 3-loop OS
vacuum-polarization coefficient $a_3^{[1]}$ of Eq.~(15) and the 4-loop OS
$\beta$-function coefficient of Eq.~(27), each given in analytical form,
for the first time. In the course of deriving $a_3^{[1]}=0.3268745\ldots$
we have devised two entirely new 3-loop programs, SLICER~\cite{bro3} and
SHELL3~\cite{tar}, which substantially validate the programs
MINCER~\cite{gor4} and RECURSOR~\cite{bro1}, respectively,
and hence confirm the large discrepancy between our result
and the numerical estimate
$a_3^{[1]} \approx1.356$ that follows from the 8th order muon-anomaly
contribution of Eq.~(20), obtained by Kinoshita~et~al.~\cite{kin1}.
In this connection, we remark that
there exists an impressive body of work~\cite{IR}
to support our belief that the methods of dimensional regularization, used
here and in Refs.~\cite{bro1,gor3,gor4}, yield final results free of any
infra-red pathology.
Our value for $a_3^{[1]}$ does not affect the result
of Ref.~\cite{kat} for the on-shell constant $b_4$, whose $N=1$ value has
recently been verified~\cite{ach1}. A formula involving $a_3$ has been
given for $b_5$~\cite{ach2}. This is in error by the omission of a term
involving the
5-loop coefficient in the expansion of Eq.~(31), whose value is unknown.

\newpage\noindent{\bf Note added: }
After completing this work, we communicated our result of Eq.~(19) to
Professor Kinoshita, who undertook a re-evaluation of the diagrams
of Fig.~1, obtaining~\cite{kpc} a numerical result substantially different
from that given in Eq.~(20) and considerably closer to ours.
\vskip1cm
\noindent{\bf Acknowledgements: }
This work is part of the recently formed ASTEC
International Project, intended to combine the efforts of scientists
working on the development and application of numerical and symbolic
manipulation computer methods in High Energy Physics and Nuclear Physics.
(For discussions of the current status of the project see
Ref.~\cite{IInd}.)~ D.J.B.\ and A.L.K.\ thank D. Perret-Gallix, one of the
main organizers of ASTEC and of its Second International Workshop on
Artificial Intelligence and Expert Systems for High Energy Physics and
Nuclear Physics, for hospitality at this workshop in L'Agelonde, France,
which initiated the collaboration resulting in this paper. O.V.T.\ is
grateful to the Physics Department of Bielefeld University for warm
hospitality. We are most grateful to Professor Kinoshita
for commenting on a preliminary version of our paper and for
undertaking a re-evaluation~\cite{kpc} of the diagrams of Fig.~1.

\end{document}